# Exploratory Testing: One Size Doesn't Fit All

Ahmad Nauman Ghazi[1], Kai Petersen[1], Elizabeth Bjarnason[2] and Per Runeson[2]

[1]Department of Software Engineering,
Blekinge Institute of Technology, Karlskrona, Sweden

[2]Department of Computer Science,
Lund University, Lund, Sweden

{nauman.ghazi, kai.petersen}@bth.se, {elizabeth.bjarnason, per.runeson}@cs.lth.se

**Abstract**

Exploratory testing (ET) is a powerful and efficient way of testing software by integrating design, execution, and analysis of tests during a testing session. ET is often contrasted with scripted testing, and seen as a choice between black and white. We pose that there are different levels of exploratory testing from fully exploratory to fully scripted and propose a scale for the degree of exploration for ET. The degree is defined through levels of ET, which correspond to the way test charters are formulated. We have evaluated the classification through focus groups at four companies and identified factors that influence the level of exploratory testing. The results show that the proposed ET levels have distinguishing characteristics and that the levels can be used as a guide to structure test charters. Our study also indicates that applying a combination of ET levels can be beneficial in achieving effective testing.

## 1. Introduction

Advocates of exploratory testing (ET) stress the benefits of providing freedom for the tester to act based on his/her skills, paired with the reduced effort for test script design and maintenance. ET can be very effective in detecting critical defects [10]. We have found that exploratory testing can be more effective in practice than traditional software testing approaches, such as scripted testing [10]. ET supports testers in learning about the system while testing [10, 12]. The ET approach also enables a tester to explore areas of the software that were overlooked while designing test cases based on system requirements [5]. However, ET does come with some shortcomings and challenges. In particular, ET can be performed in very many different ways, and thus there is no one-way of training someone to be an exploratory tester. Also, exploratory testing tends to be considered an ad-hoc way of testing and some argue that defects detected using ET are difficult to reproduce [4]. We pose that there are different levels of exploratory testing, from fully exploratory to fully scripted and propose a scale for the degree of exploration for ET.

There is very little work on providing structure and guidance for ET to utilize different degrees of exploration, despite discussions on how to benefit from ET in both industry and academia [11]. Bach introduced a technique named Session Based Test Management (SBTM) [7] that provides a basic structure and guidelines for ET using test missions. SBTM provides a strong focus on designing test charters to scope exploration to the test missions assigned to exploratory testers. A test charter provides a clear goal and scopes the test session. It can also be seen as a high level test plan [2]. However, little guidance exists of how to utilize the test missions in order to achieve different degrees of exploration.

Recently, Ghazi et al., [2] provided a checklist of contents to support test charter design in exploratory testing. The focus of that research was to support practitioners in designing test charters depending on the context of the test mission and the system under test. However, it remains to explore different degrees of exploration and how they map to the contents of the test charters, e.g. test goals, test steps, etc.

In this article, we present a classification of different levels of exploratory testing (ET), from *free style* testing to *fully scripted* testing. We exemplify the ET levels with test charter types that were defined based on studying existing test charters in industry. We have evaluated the classification and the test charter types through focus groups at four companies, Axis Communications,

Ericsson, Softhouse Consulting, and Sony Mobile Communications. The focus groups also provided insight into factors that influence the ET levels.

## 2. Exploratory testing

Exploratory testing (ET) is a way to harness the skills, knowledge and creativity of a software tester. The tester explores the system while testing, and uses the knowledge gained to decide how to continue the exploration. The design, execution, and analysis of tests take place in an integrated fashion [3]. The experience and skills of the tester plays a vital role in ET and influences the outcome of the testing [1,9]. ET displays multiple benefits, such as testing efficiency and effectiveness [10,11], a goal-focused approach to testing, a high degree of ease of use, flexibility in test design and execution, and providing an interesting and engaging task for the tester [11].

Shah et al. [4] conducted a systematic review of the literature on ET, and found that the strengths of ET are often the weaknesses of scripted testing and vice versa. They conclude that ET and scripted testing should be used together to address the weaknesses of one process with the strength of the other. While Shah et al. do not consider different types of ET, other research highlights the existence of different degrees of exploration.

## 3. Classification of levels of exploratory testing

We have identified five levels of exploratory testing (ET levels) ranging from *free style* exploratory testing, to *fully scripted* testing. Figure 1 provides an overview of the proposed classification. Each of the five levels are defined by a test charter type that guides the testing. The test charter for each ET level adds an element that sets the scope for exploration space for the tester, from the level of free style testing to fully scripted testing, which is the least exploratory level. On the freestyle level, the tester is only provided with the test object. The exploration space is reduced for each level, by adding further information to the test charter, e.g. high-level goals. The tester is thus further focused for each decreasing ET level, and the reduced freedom leads to a less exploratory approach compared to the previous level. The test charter type for the lowest ET level, i.e. fully scripted, contains both test steps and test data, and thus leaves no space for exploration during test execution.

We provide examples of test charters produced during one of the focus groups (see Box 1) in Figure 3. The test charter Figure 3for the highest ET level contains only the goal for the testing, namely to verify a specific function of the system. The test charter for the medium ET level contains additional information, e.g. suitable starting points for the testing. Finally, the test charter for the low ET level contains detailed test activities/steps in addition to goals and other information. The next level, i.e. fully scripted, which is not shown in Figure 2, also contains test data. For example, the test charter for the test activity "Copy content to card from PC" would also specify the content to be copied.

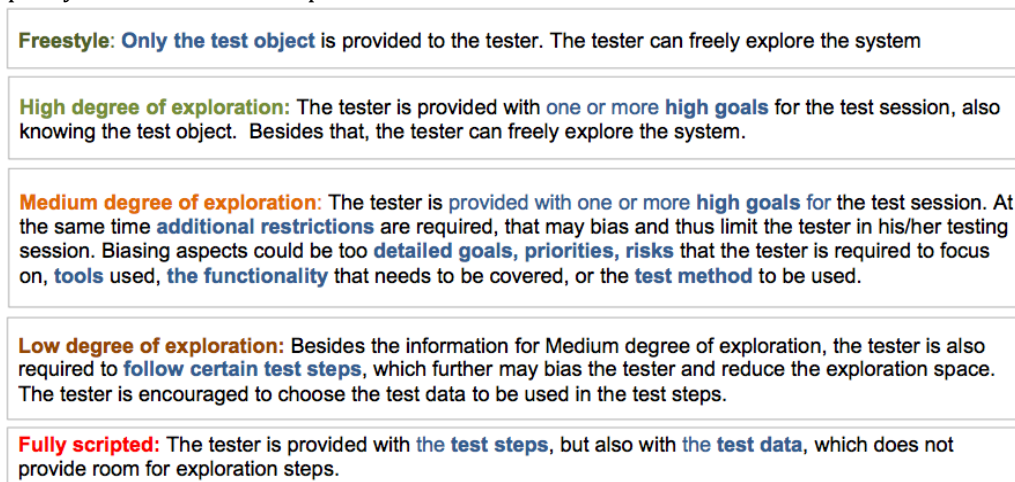

*Figure 1 - Classification of levels of exploratory testing; ET levels.*

|  | High degree | Medium degree | Low degree |
|---|---|---|---|
| Test goal/ purpose | To verify adoptable storage | 1. Test different SD cards (speed and size) 2. Test SD card and internal/external memory 3. Test move SD card to other device (phone/PC etc) 4. Test extract files from SD card | 1. To verify that adopt.storage behaves (as expected) according To the requirement 2. To test that no data loss occurs |
| Set-up (pre-conditions of what needs to be available to test) |  |  | 1. PC 2. 2 devices supported by adopt.storage 3. SD cards 1 … n 4. 1 device not supporting adopt.storage 5. Test content |
| Priority |  |  | High. There must not be any data loss |
| References |  |  | See ALMreg.doc |
| Test activities |  |  | 1. Inser SD card 2. Setup as internal 3. Copy content to card from PC 4. Read content on card from device 5. Save |
| Additional information used |  | 1. Use previously identified problems as input 2. Use Google requirements if needed |  |

*Figure 2 - Example of test charters for the high, medium and low ET levels.*

## 3. Results from the focus groups

We conducted focus groups in four companies (Axis Communications, Ericsson, Softhouse Consulting and Sony Mobile Communications) to evaluate the classification of ET levels and explore factors that influence these. We outline our research method in Box 1.

The participants of the focus groups discussed factors that influence the ET levels and the corresponding test charter types. We identified six main areas that influence the levels, namely *defect detection, time and effort, people-related factors, evolution and change, traceability* and *quality requirements*. We show these factors in Figure 3 by presenting two opposing poles for each factor. For example, better learning (indicated as positive by ☺) versus poor learning (indicated as negative by ☹). If the impact on the ET level is neutral this is indicated by 😐.

Overall, the practitioners had a positive view of the higher ET levels, i.e. freestyle and high exploration, for four of the six main areas. The participants noted a positive impact for these levels within the areas of *defect detection, time and effort, people-related factors,* and *evolution and change*. In contrast, they expressed a negative impact for factors related to *traceability* and *verifying quality requirements*. The participants believed that the more exploratory ET levels have a negative impact on these two areas.

**Defect detection:** The members of all focus groups highlighted that the exploratory approach will identify more significant defects. However, one participant stated that this may only be the case "if you know what the faults may be", i.e. the tester should have the skills to identify where significant faults are most likely to occur. Thus, the tester's skills plays a vital role in exploratory testing, as also confirmed by empirical studies on ET [1]. These skills are also required to judge whether an explored behavior is a critical defect or not. However, the participants also pointed out that people taking a new perspective often find new defects. One practitioner said: "every time we get new people in the team, we find new defects in the system". This highlights one of the benefits of exploration, namely that of not biasing the search for defects, for example, through pre-existing test cases and prior knowledge embedded in scripted tests. This may be the case for lower ET levels, i.e. low exploration and fully scripted. Some participants pointed out that high exploration comes with challenges with reproducibility of detected defect. One participant said that the "problem is when you have higher degree of exploration … the developers want to have very detailed steps to reproduce it". However, when "you focus on the reproducing you lose the exploration" which is a drawback with the fully scripted level.

**Time and effort:** Many participants highlighted time efficiency as one of the benefits of the high and medium ET levels. One practitioner explained this by saying that "we can get a better overview quickly" with higher ET levels. These higher ET levels also require less effort to prepare the test, compared to the lower and fully scripted ET levels. One participant explained that the many details of the low ET levels require "an upfront investment to develop test cases" before you can execute them. Another participant adds by saying that there is "less administration if you have a high degree of exploration because then you have quite openness and it

**Box 1: Methodology and case details:**
This article is based on the study of test processes at four companies in Sweden involved in large-scale product development in the area of telecommunications and embedded systems. The companies we studied are Axis Communications, Ericsson, Softhouse Consulting, and Sony Mobile Communications. An overview of the companies is provided below. The companies selected the focus group participants based on their experience of testing and their interest in exploratory testing.

| Company | Axis Communications | Ericsson | Softhouse Consulting | Sony Mobile Communications |
|---|---|---|---|---|
| Domain | Security cameras | Telecom | App development | Telecommunications/ mobile application development |
| Number of people in focus group | 4 | 7 | 3 | 6 |
| Development process | Agile | Agile | Agile | Agile |
| Current test focus | High degree of exploration | Mostly scripted | Mostly scripted | Mostly scripted |

We used focus groups [6] as the method for data collection at all four companies and conducted them in two main iterations. The initial two focus groups were exploratory and were conducted at two companies interested in extending their use of exploratory testing in their current test processes (Axis Communications and Sony Mobile Communications). These first two focus groups contained the following steps:
1. Introduce the basic concepts of exploratory testing
2. Present the classification of exploration levels
3. Share examples of test charters type for each level with the participants
4. The participants re-write an existing test case at the different exploration levels using the provided test charter types
5. Open discussion of how each level and test charter types matches the context for their current test practices
6. Elicit factors that affect the level of exploration in testing

We conducted the third and fourth focus groups at Ericsson and at Softhouse Consulting to validate the results from the initial focus groups. The participants were mainly experienced testers, each with more than 20 years of experience of software testing, with a strong focus on exploratory testing. We first conducted a survey with the focus group participants to gauge their views of the degree of impact of ET levels on the factors elicited from the first two focus groups (such as learning). We discussed the outcome of the survey at the (subsequent) focus group. In particular, we discussed the impact on the ET levels of exploration and reached a consensus for each factor.

We audio recorded and transcribed the focus group sessions, and analyzed these to identify resulting advantages and disadvantages of each ET level in the proposed classification.

*Threats to validity:* The participating practitioners have not experienced all the ET levels and the corresponding test charter types discussed in the focus group. However, they may relate to them given their experience of testing. We reduced this threat by letting the practitioners gain hands on experience of the test charter types during the focus group (Step 4 above). A common threat of studies with companies is the generalizability of the findings. We partially reduce this threat, by involving four companies. We mitigated the risk of research bias by involving three different researchers in performing the focus groups and in jointly discussing the outcome of these.

| Negative (☹) | Freestyle | High exploration | Medium Exploration | Low Exploration | Scripted | Positive (☺) |
|---|---|---|---|---|---|---|
| | | | Defect detection | | | |
| Finds less critical defects | ☺ | ☺ | ☺ | 😐 | ☺/☹ | Finds more significant/ critical defects |
| Does not help finding unknown defects to a great extent | ☺ | ☺ | ☺/☹ | 😐 | ☹ | Helps to uncover unknown defects |
| Difficult to reproduce defects | ☹ | 😐 | 😐 | 😐 | ☺ | Easier to reproduce defects |
| | | | Time and effort | | | |
| Time inefficient | ☺ | ☺ | 😐 | ☹ | ☹ | Time efficient |
| Effort intensive test preparation | ☺ | ☺ | 😐 | ☹ | ☹ | Less effort to prepare tests |
| More effort to maintain test cases | ☺ | ☺ | 😐 | 😐 | ☹ | Less effort to maintain test cases |
| | | | People factors | | | |
| Poor learning | ☺ | ☺ | ☺ | 😐 | 😐 | Better learning |
| Hinders critical thinking to challenge expected outcomes | ☺ | ☺ | ☺ | 😐 | ☹ | Motivates critical thinking to challenge expected outcomes |
| Demotivates the tester | ☺ | ☺ | ☺ | 😐 | ☹ | Motivates the tester |
| | | | Evolution and change | | | |
| Difficult to design new tests | ☺ | ☺ | 😐 | ☹ | ☹ | Easy to design new tests |
| Resilient to change test cases | ☺ | ☺ | ☺ | ☺/☹ | ☺/☹ | Easier/ provides freedom to change test cases |
| Difficult to fill knowledge gap when adding new requirements | ☹ | 😐 | ☺ | 😐 | ☺ | Easier to fill knowledge gap when adding new requirements |
| | | | Traceability | | | |
| Hard to trace coverage | ☹ | ☹ | ☺ | ☺ | ☺ | Easy to trace coverage |
| Inefficient in checking verification of requirements | ☹ | ☹ | ☹ | ☺ | ☺ | EEfficient in checking verification of requirements |
| | | | Quality requirements | | | |
| Difficult to verify conformance/ legal requirements | ☹ | ☹ | ☹ | ☺ | ☺ | Easier to verify conformance/ legal requirements |
| Does not help in checking performance issues | ☺/☹ | 😐 | 😐 | 😐 | ☺ | Helps to check performance issues |

*Figure 3 - Overview of factors influencing the ET levels derived from the focus groups and the survey*

is much easier to write test cases." The focus group participants indicated that the effort to maintain test cases at the higher ET levels is less since changes are more likely to affect the level of detail of test cases at the lower ET levels. For example, the tester would need to update the test steps.

**People factors:** The participants highlighted that the higher ET levels are beneficial for encouraging critical thinking, challenging the system when testing, and that they support learning. One participant said that at the highest ET level (freestyle) learning "might take longer time. But that it would probably be a better approach from the beginning to understand what the testing is." This participant also said that "you only do fully scripted when you know the system and it is monotonous and you can also get tired of it". However, some participants also expressed positive learning effects from fully scripted testing and that "it is definitely easier to start learning testing when it is fully scripted. If we do freestyle then it would be difficult" because "it requires skills and some form of competence or otherwise you are completely lost." One participant suggested that to make full use of the higher ET levels, i.e. freestyle and high exploration "you have a mentor that tells you explore this and then you explore and test. When you have questions, you go back and ask/ discuss with the mentor." Several participants pointed out that a tester's experience plays an important role and that less experienced testers are often able to identify new defects since they bring a new perspective to a project. At the same time, a tester with less experience may find it hard to conduct freestyle or high exploration testing since they do not have the domain knowledge required. Hence, there are additional factors that affect the degree of learning. Participants also pointed that with scripted testing "one problem can be that if you just keep following the test steps then there is a chance that you miss the approval criteria". Finally, the participants pointed out that learning from the requirements occurs during the derivation of the tests from the detailed requirements. They also highlighted motivation as an important distinguishing factor, where testers quickly get bored when testing at low degrees of exploration included fully scripted testing.

The participants also highlighted that the impact and effect of the ET levels may very well vary throughout the development cycle. They said that the higher ET levels might be particularly useful during the early phases of testing to explore and learn about the system. The testers may then design new tests that later become scripted tests, which are used for regression testing in later stages when the project is closer to releasing software.

**Evolution and change**: The participants highlighted that it is easier to design new tests for higher ET level (freestyle and high degree of exploration) since this requires less effort; "you have an openness and it is much easier to write test cases." In line with this, the participants also expressed that changes can be more easily implemented given that the "higher ET levels are more resistant [to change] since you don't need to change a lot of details." They also expressed that the communication around changes to tests is simplified for these higher ET levels and that when "some behavior has changed and you just discuss and notify that this has changed instead of going in details every time." However, the practitioners also said that the higher ET levels are more challenging when requirements are added or changed, since information of the new requirements is needed to guide the testing.

**Traceability**: All focus groups have highlighted that the difficulties of tracing coverage is a major drawback for higher ET levels both regarding coverage of code and of requirements. One participant said: "The sense of coverage is much lower as compared to when you ticked off 100 test cases in scripted tests." This issue also applies to requirements coverage since test cases at the higher ET levels per definition do not include any mapping to individual requirements.

**Quality requirements:** Several participants highlighted that the higher ET levels are not suitable for conformance requirements. One participant said that "We do have a lot of conformance with different standards and legal requirements" and "if you don't have this kind of [low] ET level then it is easier to miss." The participants expressed different views on this regarding performance. On the one hand, load on the system may be better generated with scripted automated tests. In one case, a participant highlighted that for performance testing "you have to continuously compare it to different firmware and we need to have similar tests again then we can't really explore a lot." However, it is also important to consider the end user perspective and make observations while testing at higher ET levels.

## 4. Conclusions and Future Work

Earlier in this article, we discussed that there have been some attempts in past to provide structure to ET and guide the test process by defining clear test missions as well as time-boxing the sessions. Our current work aims to provide practitioners a better understanding of ET practices by introducing levels of exploration. The levels of exploration that we propose have distinct elements that help the practitioners to distinguish each level clearly. This distinction of exploration levels, in turn, would facilitate the test teams to make informed decisions to more effectively test their software.

Exploratory testing can find critical and otherwise missed defects by utilizing the skill and creativity of the tester, while being insufficient for verifying conformance to requirements due to providing weak coverage of requirements. In contrast, scripted testing provides this and is a vital component in regression testing. However, the question about exploratory testing is not whether or not to apply it, but rather which levels of exploratory testing to apply to achieve the desired outcome.

We have identified five levels of exploratory testing (ET levels) from fully scripted to fully exploratory, or freestyle, and have explored factors that influence these level through a series of focus groups. Our research shows that the ET levels influence factors such as the ability to detect defects, test efficiency, learning and motivation, and that different outcomes are to be expected depending on the choice of ET level. This allows testers to select the ET level according to what they want to achieve with their testing. For example, testers operating at higher ET levels, e.g. freestyle, can expect to achieve improved defect detection, savings in time and effort, and facilitated management of evolution and change. They can also expect a positive impact with regards to learning and motivation. Though there are drawbacks too, since the higher ET levels are weak in supporting traceability and verification of quality requirements concerning conformance and performance. Another drawback with high ET levels is the weak reproducibility of defects, as the test steps are not clearly documented for developers to follow to reproduce the defect. However, we note that recent research studies provide solutions for tracking the testing session to later derive and repeat the test steps (cf. [8]).

We encourage practitioners to consider striving for a combination of exploratory and scripted testing. In this way, testers can obtain the positive effects of the higher degrees of exploration, while not neglecting other types of testing. As one participant stated at the end of the focus group: "we [now] think that we want both scripted and exploratory; a mix of both approaches, so that we can approach our testing in different ways". We also found that the test charters used to define the ET levels provided practical value to the participants. The practitioners quickly grasped the differences between the ET levels by viewing and applying these test charter types. We suggest that practitioners reflect on the ET levels by using a similar approach. They can explore and reflect on how the various ET levels could support them by rewriting existing charters or scripted tests according to the corresponding test charter types.

## Acknowledgement

We would like to thank the participating companies and in particular the individuals for their active involvement in and support of this research.

This work was partly funded by the Industrial Excellence Center EASE Embedded Applications Software Engineering, (http://ease.cs.lth.se).